\documentclass[aps,amsfonts,pra,twocolumn,showpacs,eqsecnum]{revtex4-1}
\usepackage{epsfig,amsmath,amssymb,bm,epsf,graphics,psfrag,verbatim,subfigure}
\usepackage{slashed}
\def\confenergy{U} % The energy of a configuration of the polymer system
\def\plength{L} % The longitudinal length of the polymer system
\def\pwidth{w} % the width of the polymer *and* the quantum system
\def\qtime{t} % the (imaginary) time coordinate in the quantum picture
\def\qham{H} % quantum Hamiltonian

\begin{document}

\title{Directed-polymer systems explored via their quantum analogs:\\ General polymer interactions and
their consequences}
\author{D. Zeb Rocklin$^{1}$ and Paul M. Goldbart$^{2}$}

\affiliation{
$^1$Department of Physics, University of Illinois at Urbana-Champaign,
1110 West Green Street, Urbana, Illinois 61801, USA}

\affiliation{
$^2$School of Physics, Georgia Institute of Technology,
Atlanta, Georgia 30332, USA}

\date{\today}

\begin{abstract}

The impact of polymer-polymer interactions of various types on the thermodynamics, structure, and accommodation of topological constraints is addressed for systems comprising many directed polymers in two spatial dimensions.
The approach is predicated on the well-known equivalence between the classical equilibrium statistical mechanics of directed polymers in two spatial dimensions and the imaginary-time quantum dynamics of particles in one spatial dimension, originally exploited by P.-G.~de~Gennes [J.\ Chem.\ Phys.\ {\bf 48\/}, 2257 (1968)].
Known results concerning two exactly solvable microscopic models of quantum particles moving in one spatial dimension---the Lieb-Liniger model of contact interactions and the Calogero-Sutherland model of long-range interactions---are used to shed light on the behavior of the corresponding polymeric systems.
In addition, the technique of bosonization is used to reveal how generic polymer interactions give rise to an emergent polymer fluid that has universal collective excitations.
Comparison of the response to topological constraints of a fluid of simply noncrossing (i.e., noncrossing but otherwise noninteracting) directed polymers, explored in a companion Paper, to the response of a generically interacting directed polymer fluid reveals that the structure is quantitatively unchanged by the generic interactions on the line transverse to the pin, and is qualitatively unchanged by the generic interactions throughout the two dimensions of the system's extent.
That is to say, in response to a topological pin constraint there is a divergent pile-up in polymer density on the compressed side of the pin and a gap of finite area, within which the polymer density is negligibly small, on the other side of the pin.
Furthermore, the free-energy cost associated with a pin that partitions a system having generic interactions is found to be proportional to the pin-partitioning cost for a system of simply noncrossing polymers.
\end{abstract}

\maketitle

\section{Introduction}

In two-dimensional settings, the statistical-mechanical properties of macroscopic systems composed of long, flexible, directed polymers are determined largely by the interactions of the polymers with one another.  Due to the reduced dimensionality, polymers encounter one another with enhanced probability as they undergo thermal fluctuations in their conformations.  As a result, even numerically weak interactions suffice to give rise to an emergent polymer fluid having physical characteristics that are essentially collective, rather than being primarily of the single-polymer type.  In the present Paper, we explore how the form and strength of interactions modifies the statistical properties of the polymer fluid.  We do this by making use of a familiar analogy between the classical equilibrium statistical physics of directed-polymer systems and the quantum physics of particle systems, according to which the thermal fluctuations of the polymer system are the analogs of the quantum fluctuations of the particle systems.

The ensemble of configurations of a set of directed, linelike objects embedded in $(D+1)$-dimensional space can be reinterpreted as the ensemble of worldlines of a corresponding set of quantum-mechanical point particles evolving in time in $D$ spatial dimensions.  A well-known mapping then relates the classical equilibrium statistical mechanics of the set of directed one-dimensional objects in the canonical ensemble to the imaginary-time evolution of the state of the corresponding set of quantum particles.  De~Gennes~\cite{pdg} first introduced and exploited this mapping in order to describe the equilibrium structure of directed noncrossing polymers confined to two dimensions, thus providing a scheme that accounts nonperturbatively for the strong local polymer-polymer interactions that serve to prevent polymer crossings.

In a companion Paper~\cite{rocklin}, hereafter referred to as~I, we considered the statistical mechanics of such a system of noncrossing polymers, but whose accessible configurations were subject to one of various topological constraints. Here, we consider systems of fluctuating linelike objects that are subject to more general classes of interactions, and address the response of these systems to topological constraints on their configurations.  In particular, we consider polymer systems that have finitely (rather than infinitely) strong contact repulsion, so that crossings of the polymers do occur but with energetic penalty, and we furthermore allow the systems to have certain specific forms of long-range polymer-polymer interactions, in part chosen for their tractability.  Additionally, we use the many-body technique of bosonization to analyze systems of polymer that are subject to a generic interaction.

The systems under consideration are all amenable to the general framework discussed in detail in~I, in which a system configuration, given by a set of polymer profiles $\{x_n(\tau)\}_{n=1}^N$, has configurational energy
\begin{eqnarray}
&&\confenergy\left[\{x_n(\cdot)\}\right]=
\frac{A}{2}\sum_{n=1}^{N}
\int_0^{\plength} d\tau\,
\big(
\partial_\tau x_{n}(\tau)
\big)^2
\nonumber
\\
&&\qquad
+\frac{1}{\plength}\sum_{n=1}^{N}
\int_0^{\plength} d\tau\,
\Phi\big(x_{n}(\tau)\big)
\nonumber
\\
&&\qquad
+\frac{1}{\plength}
\sum_{1\le n<n^{\prime}\le N}
\int_0^{\plength} d\tau\,
V\big(x_{n}(\tau)-x_{n^{\prime}}(\tau)\big),
\end{eqnarray}
\noindent
in which $A$ is the effective polymer tension, $\Phi(\cdot)$ is a one-body potential, and $V(\cdot)$ is the two-body interaction between polymers~\cite{tlocal}.  The system has length $L$ in the longitudinal direction ($\tau$) preferred by the polymers and width $w$ in the lateral direction ($x$).

\begin{figure}[hh]
\centerline{
\includegraphics[width=.5\textwidth]{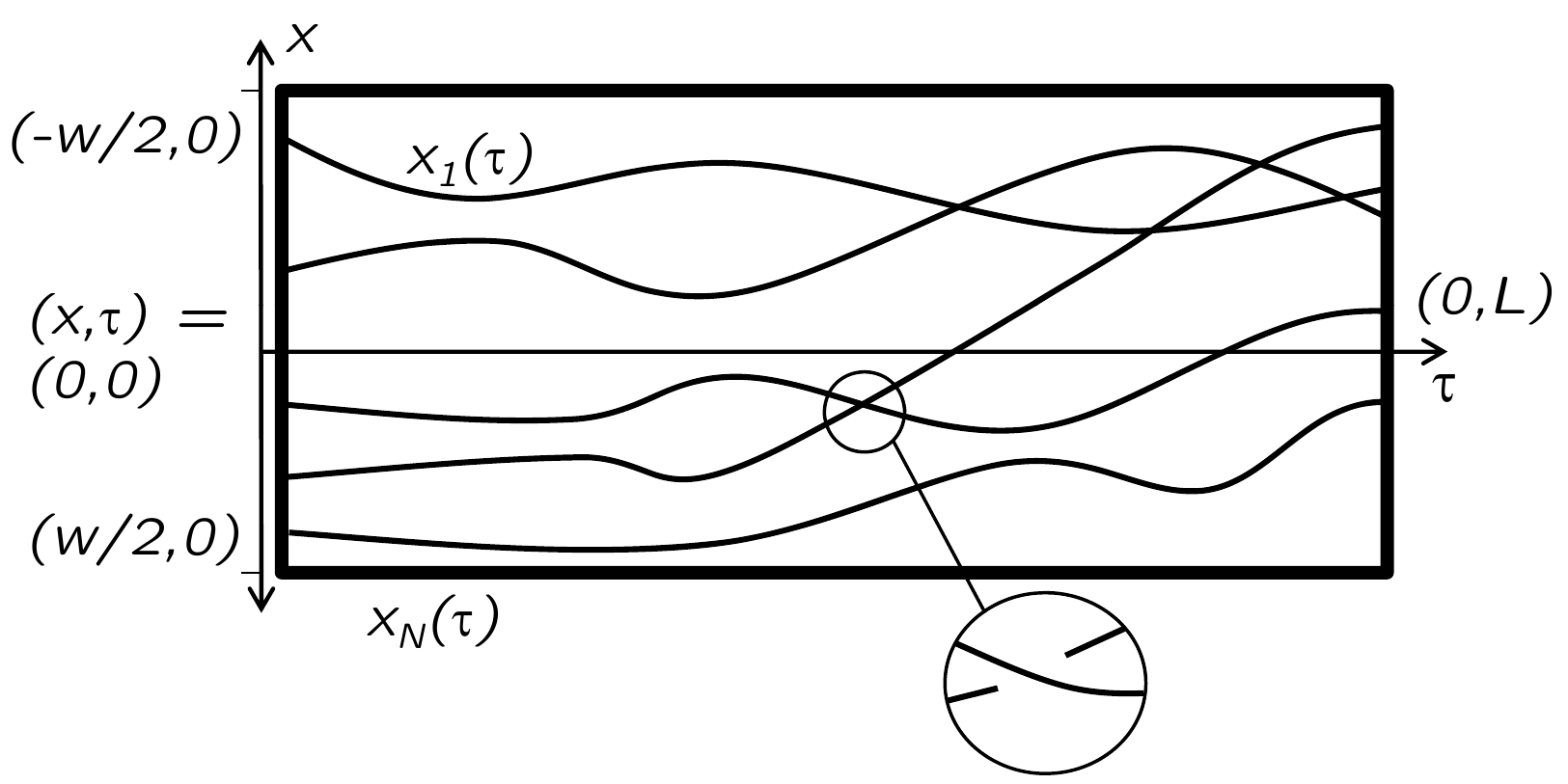}}
\caption{
The paths $\{x_n(\tau)\}$ describe a possible configuration of the directed polymer system.  Thermal fluctuations permit the system to adopt energetically disfavored configurations.  When polymers appear to intersect in the $(x,\tau)$ plane, in reality one crosses over the other by exploiting the presence of a third dimension.
(This figure originally appeared in~I.)
}
\label{fig:crossingdiagram}
\end{figure}

In addition to polymer systems, a number of other statistical systems can effectively be modeled as consisting of interacting, directed, linelike degrees of freedom.  For example, step edges on miscut crystalline surfaces~\cite{twd}, vortex lines in planar Type-II superconductors~\cite{nelson, devereaux, mpafisher, kafri, radzihovsky}, and the growing interfaces of the Kardar-Parisi-Zhang universality class~\cite{kulkarni} all fall within the framework discussed here, and have all previously been studied via their quantum analogs.

The present Paper is organized as follows.
In Sec.~\ref{sec:cp} we consider a system of crossing polymers having a finitely strong contact repulsion.
In Sec.~\ref{sec:lr} we consider a system of noncrossing polymers that interact via a long-range interaction.
In Sec.~\ref{sec:bosonization} we use bosonization to describe a polymer system subject to generic interactions.
In Sec.~\ref{sec:pin} we consider the effect of topological constraints on the polymer systems considered in the two prior sections.
In Sec.~\ref{sec:conclusion} we summarize our results and give some concluding remarks.

\section{Crossing polymers}
\label{sec:cp}

In a previous Paper, I, we enforced the noncrossing condition $x_n(\tau) \neq x_{n'}(\tau)$ for $n \neq n'$ by including an infinitely strong repulsive contact potential.
Let us consider now a system in which the contact repulsion between polymers is finite, so that polymer lines may cross one another, albeit with an energetic penalty:

\begin{eqnarray}
&&
\confenergy\left[\{x_n(\cdot)\}\right]=
\frac{A}{2}\sum_{n=1}^{N}
\int_0^{\plength} d\tau\,
\big(
\partial_\tau x_{n}(\tau)
\big)^2
\nonumber
\\
&&
\qquad
+ 2 c \sum_{1\le n<n^{\prime}\le N}
\int_0^{\plength} d\tau\,
\delta\big(x_{n}(\tau)-x_{n^{\prime}}(\tau)\big),
\end{eqnarray}

\noindent where $\delta(\cdot)$ is the one-dimensional Dirac delta function and $c$ ($>0$), which has units of energy, describes the effective contact repulsion between polymers.

\begin{widetext}

\begin{figure}[hh]
\mbox{
\subfigure{\includegraphics[width=.52\textwidth]{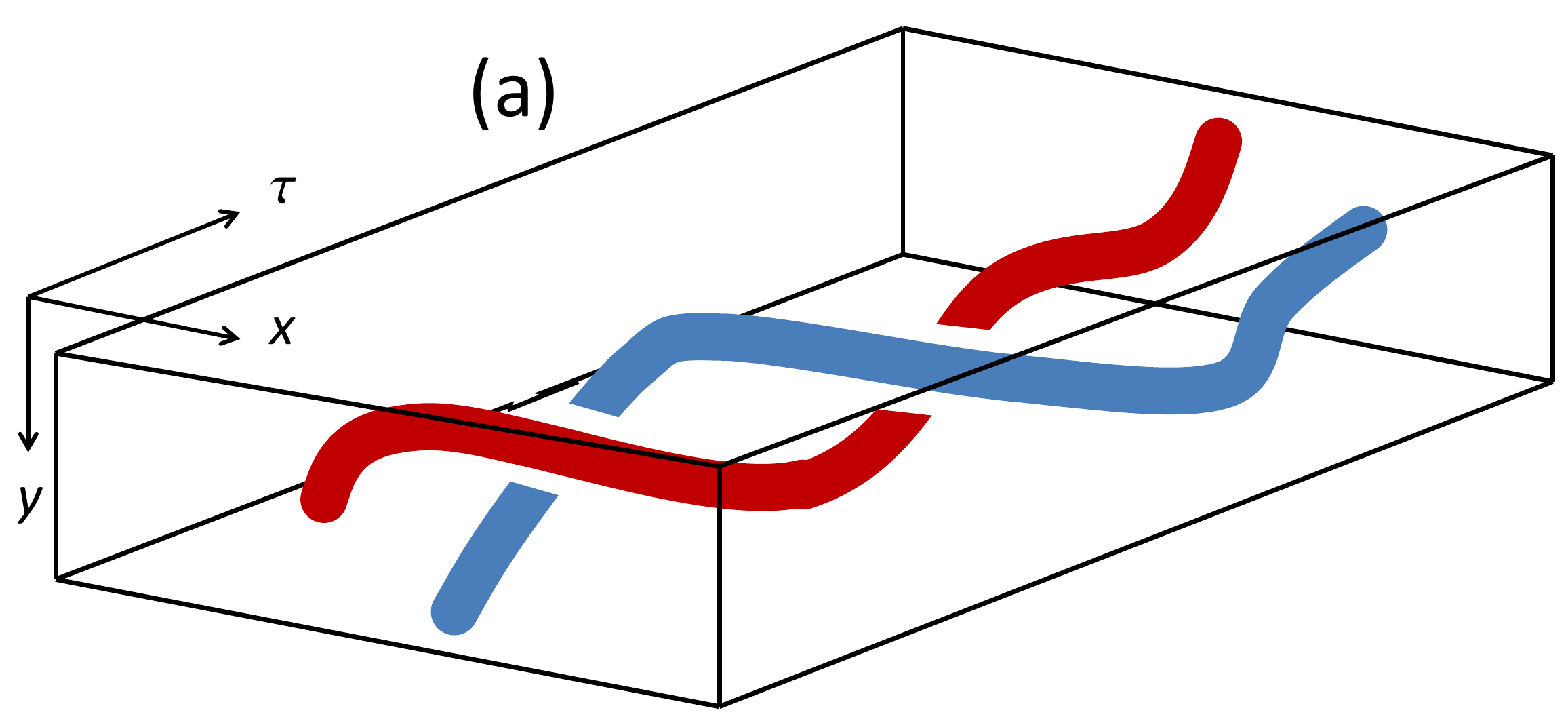}}
\quad
\subfigure{\includegraphics[width=.42\textwidth]{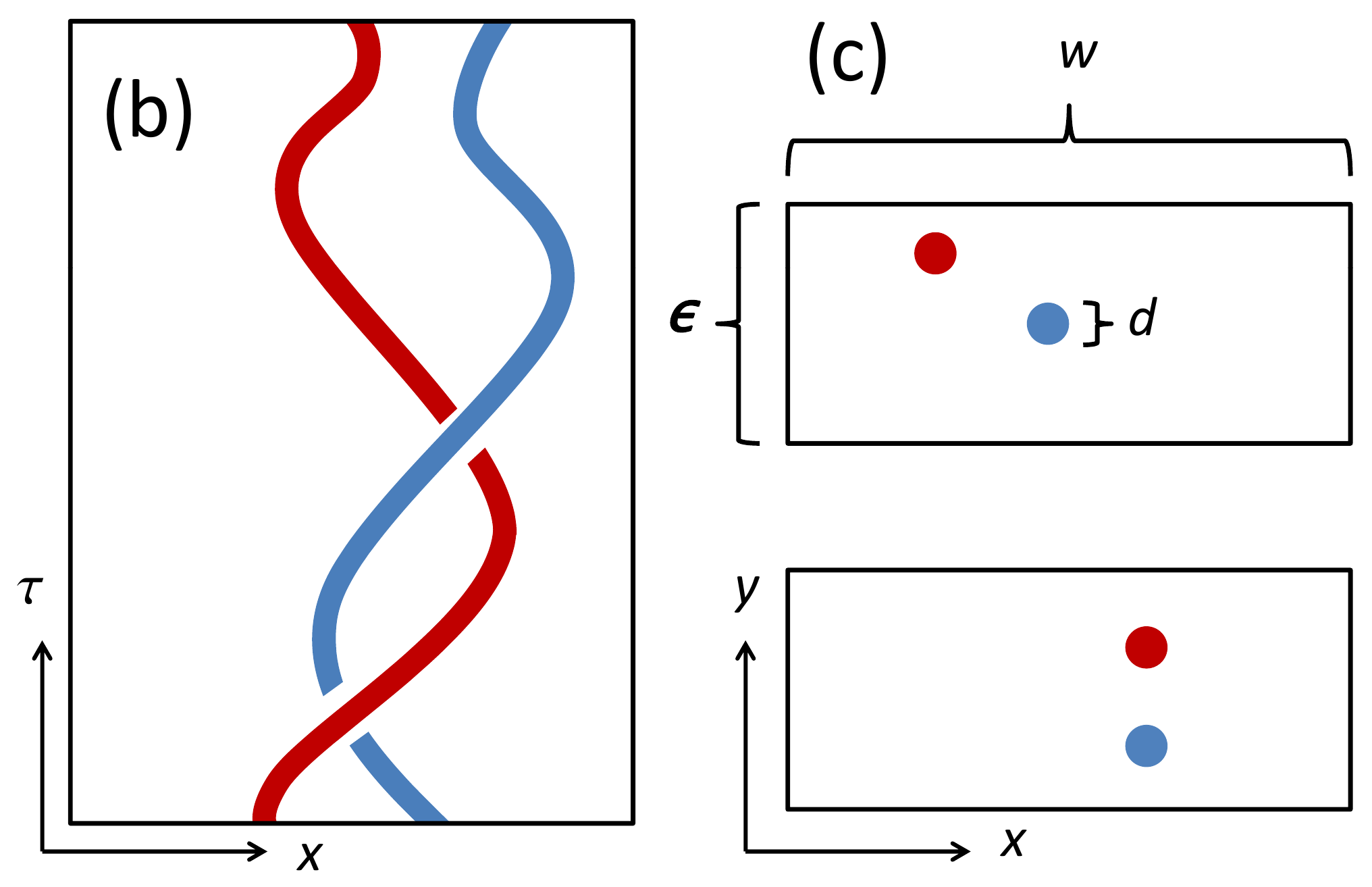}}
}
\caption{\label{phase}
(a)~A  configuration of two directed lines of finite thickness $d$ less than the thickness $\epsilon$ of the system in the $y$ direction. The polymers are envisioned to interact via the excluded-volume effect. (b)~The projection of the configuration shown in (a) onto the $x-\tau$ plane. Because the width of the system in the $x$ direction is much greater than its thickness in the $y$-direction, we may neglect the additional tension energy associated with deflections in the $y$ direction.  (c)  Cross-sections of the configuration in the plane transverse to the preferred direction of the polymers. Because of the excluded volume effect, there are fewer ways to vary the $y$ coordinates of the polymer segments in the configuration in the bottom panel than in the configuration in the top panel. This leads to a finite effective entropic repulsion between polymer configurations in the $x-\tau$ plane. (Color online)
}
\label{fig:crossingdiagram}
\end{figure}

\end{widetext}

\subsection{Origin of contact repulsion}

Real polymer systems exist in three dimensions. Let us therefore parametrize the polymer configurations by paths through three-dimensional space: $\mathbf{r}(\tau) \equiv x(\tau)\hat{\mathbf{x}} + y(\tau)\hat{\mathbf{y}}$, where $\{\hat{\mathbf{x}},\hat{\mathbf{y}},\hat{\mathbf{\tau}}\}$ form an orthogonal triad. The polymers are confined by hard walls so that $x(\tau) \in \left(-\pwidth/2,\pwidth/2\right)$ and $y(\tau) \in \left(-\epsilon/2,\epsilon/2\right)$, with $\epsilon \ll \pwidth$. Suppose further that polymers have an effective diameter $d$ such that the ensemble permits only configurations for which, for all $n,n'$ and at all $\tau$, we have

\begin{eqnarray}
\left|\mathbf{r}_n(\tau) - \mathbf{r}_{n'}(\tau)\right|^2 \geq d^2.
\end{eqnarray}

We wish to integrate out the small third dimension and restore our effectively two-dimensional picture of polymer configurations. In particular, let us integrate out two polymer coordinates $y_n$ and $y_{n'}$ at some particular value of $\tau$. If $x_n \not \approx x_{n'}$, each of the two polymer coordinates may occupy any point in $\left(-\epsilon/2,\epsilon/2\right)$ [ignoring edge effects at $y = \pm \epsilon/2$]. In contrast, if $x_n \approx x_{n'}$ then any value of $y_n$ restricts $y_{n'}$ to a length $\epsilon-d$ in the $y$ direction. Thus, upon integrating out the third dimension we ought to reduce the weight of a polymer configuration $\{x_n(\cdot)\}$ by one factor of 

\begin{eqnarray}
\label{eq:crossingfactor}
\frac{\epsilon(\epsilon-d)}{\epsilon^2}
\end{eqnarray}

\noindent for each time one polymer crosses another. We may take the number of polymer crossings to be proportional to

\begin{eqnarray}
\sum_{n<n'} \int_0^{\plength} d\tau\,
\delta\big(x_{n}(\tau)-x_{n^{\prime}}(\tau)\big),
\end{eqnarray}

\noindent provided that, as discussed in Appendix~\ref{sec:ell}, we assume some short-distance cutoff length-scale $\ell$ such that $\dot{x}(\tau)$ remains finite at a crossing. Then we obtain an effective contact repulsion from this entropic effect, with parameter

\begin{eqnarray}
2 c= -T \ln(1-d/\epsilon),
\end{eqnarray}

\noindent where $T$ denotes the temperature in units of energy~\cite{cutoff}. Note that, in the limit in which polymer width approaches the width of the system in the third dimension, the condition that no polymer crossings occur is recovered.

\subsection{Quantum analog}

The analog of the polymer system subject to a contact repulsion is a one-dimensional gas of Bose particles interacting via a contact repulsion. Such a system is governed by the Lieb-Liniger Hamiltonian $\hat{H}$, given by

\begin{eqnarray}
\hat{H} = -\frac{\hbar^2}{2 m} \sum_{n} \frac{\partial}{\partial x_n} + 2 c \sum_{n<n'} \delta\left(x_n - x_{n'}\right),
\end{eqnarray}

\noindent where $n = 1, \ldots , N$ labels the particles.
This describes an integrable system, with each eigenstate of the Hamiltonian being characterized by a set of $N$ quasi-momenta $\{ k_i \}$ (see Ref.~\cite{ll}). In the limit $c \rightarrow +\infty$, the system may be mapped exactly onto a system of free fermions.  For finite values of $c$, the quasi-momenta lie in some band (for periodic boundary conditions) $k_i \in (-K,K)$, where $K$ is less than or equal to the Fermi momentum of the corresponding system of free fermions. This means that the logarithmic divergence in the X-ray form factor, found for a system of simply noncrossing polymers by de Gennes at wave number $k=2 \pi N/w$~\cite{pdg}, will instead occur at some lesser value that is dependent on the interaction strength $c$.

As discussed in I, the free energy-density $\mathcal{F}/w L$ of the long polymer system may be obtained from the ground-state energy of the quantum system; the result is

\begin{eqnarray}
\frac{\mathcal{F}}{w L} = \frac{1}{2} \left(\frac{N}{w}\right)^3 \frac{1}{\beta^2 A} e(\gamma)
\end{eqnarray}

\noindent where $e(\cdot)$ is the dimensionless energy function that Lieb and Liniger derive by means of the Bethe Ansatz. In terms of polymer parameters, the dimensionless interaction parameter $\gamma$ is given by

\begin{subequations}
\begin{eqnarray}
\gamma &=& 2 c \left(\frac{w}{N}\right) \beta^2 A \\
&=&-\left(\frac{w}{N}\right) A \beta \ln \left(1-d/\epsilon\right)
.
\end{eqnarray}
\end{subequations}

\begin{figure}[hh]
\centerline{
\includegraphics[width=.5\textwidth]{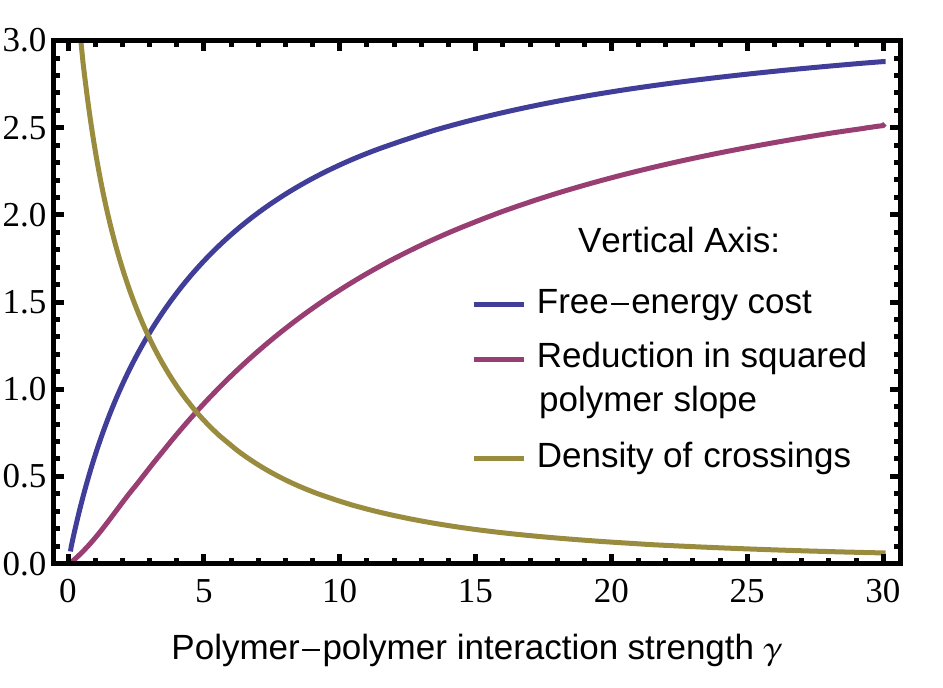}}
\caption{
Basic (scaled) characteristic quantities of the polymer system, as a function of the dimensionless parameter $\gamma$ that describes the strength of polymer-polymer interactions. Both the free-energy density (blue line) and the reduction in the squared polymer slope $\langle \dot{x}^2(\tau) \rangle$ (red line) monotonically approach their
values for simply noncrossing polymers as $\gamma$ is increased. The yellow line, depicting the areal density of polymer crossings, diverges for small $\gamma$ and approaches its own simply noncrossing value (i.e., zero) for large $\gamma$. (Color online).
}
\label{fig:llplot}
\end{figure}

\noindent In the limit $\gamma \rightarrow +\infty$, one has $e(\gamma) \rightarrow \pi^2/3$, which correctly reproduces the result for simply noncrossing polymers. For $c=0$, the polymers behave independently and the $O(N^3)$ contribution to the free energy vanishes. Note that $\gamma \rightarrow +\infty$ is \emph{not} the high-density limit. Although the free-energy density increases with increasing polymer density, increasing polymer density \emph{decreases} $\gamma$, indicating that the free energy of the high-density system depends more on the polymer deflections than the polymer crossings. Given the entropic origin of the contact repulsion [cf. Eq.~(\ref{eq:crossingfactor})], we also have that $\gamma$ is \emph{inversely} proportional to temperature. At low temperatures, the polymers are frozen into a configuration having few crossings, whereas for high temperatures they cross more freely, although still less often than truly noninteracting polymers would.
The free energy is shown in  Fig.~\ref{fig:llplot}. For a sufficiently long system, any inter-polymer interaction, no matter how weak or short-ranged, suffices to  modify strongly the behavior of the system as compared to a system of free polymers. As we saw in I for simply noncrossing polymers, the physics of polymers that are subject to generic interactions is profoundly different in reduced dimensions from that of free polymers.

For a long system we can readily obtain the areal density of inter-polymer crossings, i.e.,

\begin{eqnarray}
\label{eq:crossingdensity}
\frac{1}{w L} \Bigg \langle \sum_{n<n'} \int d \tau \, \delta\left(x_n - x_{n'}\right) \Bigg \rangle = \hspace{1cm}
\nonumber \\
 -\frac{1}{2 \beta} \frac{1}{w L} \frac{\partial}{\partial c} \ln \mathcal{Z}
=
\frac{1}{2} \left( \frac{N}{w} \right)^2 e'(\gamma).
\end{eqnarray}

\noindent In the limit of strong interactions, the number of polymer crossings drops to zero, recovering the simply noncrossing system considered in I.

One may also obtain the mean squared polymer slope for the interacting polymer system:

\begin{eqnarray}
\langle \dot{x}^2(\tau) \rangle
= -\frac{2}{ \beta} \frac{1}{N L} \partial_A \ln \mathcal{Z} = \hspace{2cm}
\nonumber \\
\frac{1}{A \beta \ell}-\frac{1}{\beta^2 A^2}\left(\frac{N}{w}\right)^2 \left[e(\gamma)-\gamma e'(\gamma)\right].
\end{eqnarray}

\noindent The first term on the right-hand side is inversely proportional to the cutoff length $\ell$, as discussed in Appendix~\ref{sec:ell}. The interactions strictly decrease the average polymer slope, as greater slopes result in more polymer crossings and thus receive energetic penalties.

One may also obtain the pressure on the walls containing the polymers at $x = \pm w/2$ and $y = \pm \epsilon/2$. In the $x$-direction, the pressure $P_x$ is given by

\begin{eqnarray}
P_x = \frac{1}{\epsilon L} \frac{1}{\beta}\partial_w \ln Z = \left(\frac{N}{w}\right)^3 \frac{1}{\beta^2 A \epsilon} e(\gamma).
\end{eqnarray}

\noindent In the $y$-direction, on the other hand, the effect of interactions is to generate a pressure $P_y$ given by

\begin{eqnarray}
P_y =  \frac{1}{w L} \frac{1}{\beta}\partial_\epsilon \ln Z =
 \hspace{2.5cm}
\nonumber \\
\frac{1}{\beta} N \frac{L}{\ell}
+\frac{1}{2\beta}\left(\frac{N}{w}\right)^2 e'(\gamma) \frac{d}{\epsilon(\epsilon-d)}.
\end{eqnarray}

\noindent The first term on the right-hand side is simply the one associated with confining $N L/\ell$ polymer segments to a finite thickness (see Appendix~\ref{sec:ell}). The second term is the expected number of polymer crossings, from Eq.~(\ref{eq:crossingdensity}), multiplied by the entropic cost of ensuring that the polymers pass around rather than through one another at each crossing.

\section{Long-range interactions}
\label{sec:lr}

In the previous section we considered an interaction that was softer than the noncrossing condition considered in I. Now we wish to consider an interaction between polymers that is more powerful--- in the sense that polymers repel (or attract) one another at a distance.

\subsection{Finite-diameter effects}

Previously, we considered polymers of finite diameter that were narrow enough to cross over one another, i.e., $d<\epsilon$. Alternatively, one could return to the case of noncrossing polymers, i.e., $d=\epsilon$, whilst still taking into account the finite thickness of the polymers. In this case, the noncrossing condition becomes the requirement that no polymer may come within a distance $d$ of another, or a distance $d/2$ of the walls of the system:
\begin{eqnarray}
-w/2 < x_1 - d/2 < \ldots \hspace{3cm}
\nonumber \\
< x_j - d(j-1/2) < \ldots < (w/2)-N d.
\end{eqnarray}

\noindent This issue is readily addressed by mapping the problem onto one of simply noncrossing zero-diameter polymers via $x_j' =x_j - d(j-1/2)$. Then, the finite-diameter polymer system is equivalent to a zero diameter one having a system width narrower by $N d$. This leads to the free energy density

\begin{eqnarray}
\frac{\mathcal{F}}{w L}= \frac{\pi^2}{6} \frac{N^3}{w \left(w-N d \right)^2} \frac{1}{\beta^2 A}.
\end{eqnarray}

\begin{widetext}

\subsection{Power-law repulsion}

Next, we wish to consider a form of long-range effective interaction between polymers that, as with the simply noncrossing case,  corresponds to an integrable and exactly solvable quantum system. Such an interaction is given by

\begin{eqnarray}
\label{eq:invsquare}
V_0(x_{n}-x_{n'}) = \frac{1}{2 A \beta^2} \frac{\lambda(\lambda-1)}{(x_{n}-x_{n'})^2}.
\end{eqnarray}

\noindent Here, the parameter $\lambda$ ($>0$) gives the strength of the interaction, either attractive or repulsive. For the attractive case ($\lambda<1$), we nevertheless retain the noncrossing condition by assuming some additional short-range repulsion of sufficient strength.
Electrical dipole moments, if present on the lines, would lead to an interaction of this form,  as has been noted in the context of crystalline step edges (where the edges between the crystal steps are the thermal linelike degrees of freedom)~\cite{jaya}. In addition, for the crystalline step edges the elasticity of the crystal  gives rise to an effective repulsive interaction between step edges of this long-range form~\cite{marchenko}.
 The form of the interaction in Eq.~(\ref{eq:invsquare}) is suitable for a system of infinite width. For a system having periodic lateral boundary conditions, one must include image terms for $x_n$ at $x_n \pm w, x_n \pm 2w, \ldots$. In the presence of hard walls, as we consider here, images must also be included at $-x_n \pm w, -x_n \pm 2w, \ldots$. Then, using the mathematical identity (see, e.g., Ref.~\cite{hofbauer})

\begin{eqnarray}
\label{eq:mident}
\sum_{j=-\infty}^{j=\infty} \frac{1}{\left(j+a\right)^2} = \frac{\pi^2}{\sin^2 \left( \pi a \right) },
\end{eqnarray}

\noindent one has the polymer interaction appropriate to the finite system with hard walls at $x=\pm w/2$:

\begin{eqnarray}
V(x_i-x_j) = \frac{1}{2 A \beta^2} \left(\frac{\pi}{w}\right)^2 \left(\frac{\lambda(\lambda-1)}{\sin \left[ \pi(x_{n}-x_{n'})/w\right]^2}+
\frac{\lambda(\lambda-1)}{\sin \left[ \pi(x_{n'}+x_{n'})/w\right]^2}\right).
\end{eqnarray}

\noindent The corresponding quantum system is known as the Calogero-Sutherland model~\cite{calogero, sutherland}, and is exactly solvable. In particular, the ground-state wave function is, for a system with hard-wall boundary conditions~\cite{frahm},

\begin{eqnarray}
\label{eq:gswavefcncs}
\psi_{\mbox{gs}}(\{x_n\})
=
\bigg(
\prod_{n=1}^{N}
\big \vert \cos \frac{ \pi x_{n}}{\pwidth} \big \vert ^\lambda
\bigg) \hspace{2cm}
\nonumber \\
\times 
\bigg(
\prod_{1\le n<n^{\prime} \le N}
\!\!\!\!\!\!
\big\vert\sin \frac{\pi x_{n}}{\pwidth}-
\sin \frac{\pi x_{n^\prime}}{\pwidth}
\big\vert^\lambda
\bigg),
\end{eqnarray}

\noindent and the cost in polymer system free-energy density of the interaction, as derived from the ground-state energy of the quantum system, is

\begin{eqnarray}
\frac{\mathcal{F}}{w L} = \frac{\pi^2}{6} \left(\frac{N}{w}\right)^3 \frac{1}{\beta^2 A} \, \lambda^2.
\end{eqnarray}

\end{widetext}

\begin{figure}[hh]
\label{fig:lambdaplots}
\centerline{
\includegraphics[width=.5\textwidth]{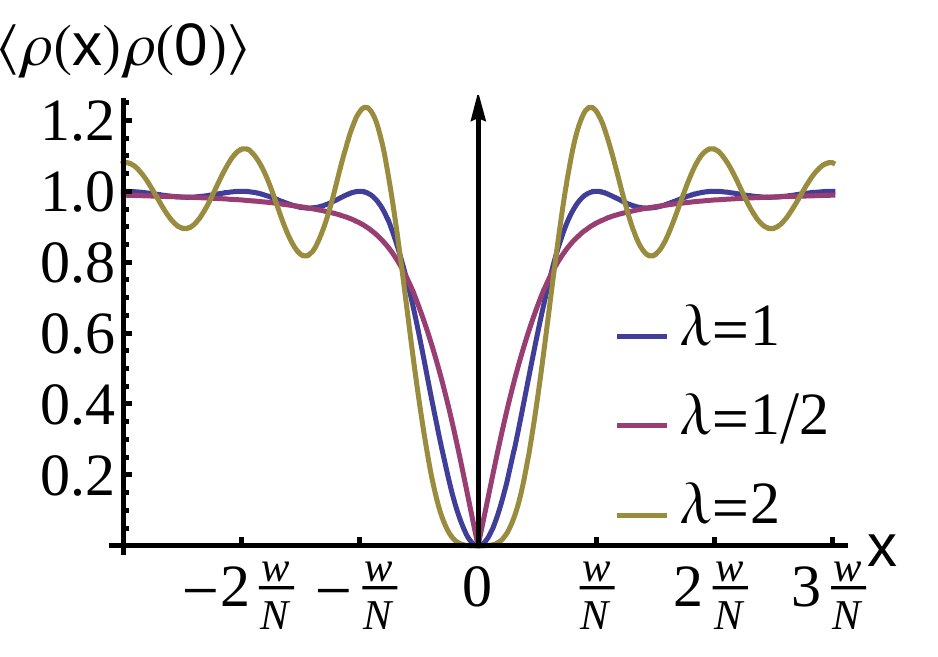}}
\caption{
Density correlations $\langle \rho(x)\rho(0) \rangle$ \big [in units of the squared average density $(N/w)^2$\big ], as a function of the lateral separation 
\big[in units of the average polymer spacing $(w/N)$\big]
 for various values of the interaction parameter $\lambda$. For $\lambda=1/2$ (i.e., attractive long-range interactions) there is no oscillatory behavior. For $\lambda=1$ and $\lambda=2$, corresponding respectively to no long-range interaction and to long-range repulsion, there are oscillations in the correlations, in the latter case leading to negative correlations for $x \approx \pm w/n, \pm 2 w/n, \ldots$. In all cases, the correlations decay as $x^{-2}$. (Color online).
}
\label{fig:lambdaplots}
\end{figure}

This model admits both repulsive ($\lambda>1$) and attractive ($0<\lambda<1$) regimes, provided that the polymers nevertheless retain a short-range repulsion that prevents them from crossing. In addition to the free-energy density, Other thermodynamic quantities, such as the reduction in the mean squared polymer slope and the pressure on the system walls, also generally scale as $\lambda^2$.

Via the quantum analogy, we may also use the ground-state wave function to relate polymer density correlations to quantum correlators. In particular, the polymer density-density correlator (in an infinitely long system) is given by

\begin{eqnarray}
\langle \rho(x) \rho(0) \rangle = \hspace{6cm}
\\ \nonumber
\int d x_3 \, dx_4 \, \ldots dx_n \, \left| \psi_{\mbox{gs}}\left(x,0,x_3,x_4,\ldots,x_N\right)\right|^2.
\end{eqnarray}

\noindent For certain values of $\lambda$, this correlation may be obtained exactly; the results are shown in Fig.~\ref{fig:lambdaplots}. The random matrix theory results used to obtain this correlator, and their application to the Calogero-Sutherland system, are discussed by Sutherland in Ref.~\cite{sutherland2}.

As suggested by Fig.~\ref{fig:lambdaplots}, for all values of $\lambda$ the probability of two particles coinciding vanishes. The noncrossing condition, along with thermal fluctuations, overwhelms even attractive long-range interactions. For attractive interactions, no oscillatory behavior occurs. For long-range repulsion or for no long-range interactions, oscillations in the density correlation, of period $w/N$, are present. In the former case, near separations $x =\pm w/N, \pm 2 w/N, \ldots$ polymers are actually more likely to be found than one would have for noninteracting (i.e., freely crossing) polymers.

\section{Bosonization}
\label{sec:bosonization}

Bosonization is a powerful technique in the study of one-dimensional quantum many-body systems. In it, the microscopic degrees of freedom are eliminated and the low-energy behavior of the system is characterized by its dominant, long length-scale, density fluctuations. By means of the mapping between polymer lines and quantum particles, bosonization can also be used to analyze the behavior of systems of interacting directed polymers in two dimensions~\cite{firstlinebose}.

The bosonization technique applies generally to quantum Hamiltonians of the form
\begin{eqnarray}
\label{eq:qhgen}
H= \int d x \, \frac{\hbar^2}{2 m} \left(\partial_x \Psi^{\dagger}(x)\right)\left(\partial_x \Psi(x)\right) \hspace{2cm}
\\ \nonumber
+ \frac{1}{2}\int \int d x \, d x' \,
\Psi^\dagger(x) \Psi(x) V(x-x') \Psi^\dagger(x') \Psi(x'),
\end{eqnarray}

\noindent where $\Psi^\dagger(x)$ represents either a bosonic or fermionic particle creation operator. For a full description of bosonization, see, e.g., Giamarchi
~\cite{giamarchi}, whose notation we adopt. The original bosonic creation operator may be expressed as

\begin{eqnarray}
\label{eq:phitheta}
\Psi_B^{\dagger}(x) = \left[\frac{N}{w}-\frac{1}{\pi} \frac{\partial \phi}{\partial x}
\right]^{1/2} \hspace{3cm}
\nonumber \\
\times
\sum_{p=-\infty}^{\infty} e^{2 i p \left(N \pi x /w - \phi(x) \right)} \, e^{-i \theta(x)},
\end{eqnarray}

\noindent in terms of the emergent fields $\phi(x)$ and $\theta(x)$, which respectively correspond to changes in the polymer density and slope (or, in the quantum case, to the amplitude and phase of the one-particle creation operator). These emergent fields obey the bosonic canonical commutation relations

\begin{eqnarray}
\left[ \partial_x \phi(x), \partial_{x'} \theta(x') \right] = i \pi \delta ' (x - x'),
\end{eqnarray}

\noindent where $\delta'(\cdot)$ is the first derivative of the Dirac delta function. The polymer density takes the form

\begin{eqnarray}
\label{eq:phidensity}
\rho(x,\tau) = \left[\left(N/w\right)-\left(\partial_x \phi(x)/\pi\right)\right] \hspace{2cm}
\nonumber \\
\times
\sum_{p=-\infty}^{\infty} e^{2 i p \left(N \pi x /w - \phi(x,\tau) \right)},
\end{eqnarray}

\noindent where the $p \neq 0$ terms correspond to the short-distance polymer structure.

The benefit of this bosonization procedure is that for a wide variety of interactions in Eq.~(\ref{eq:qhgen}) the Hamiltonian governing the low-energy excitations may be expressed in terms of the emergent bosonic fields as

\begin{eqnarray}
H = \frac{1}{2 \pi} \int d x \, \left[\frac{u K}{\hbar} \left( \partial_x \theta(x) \right)^2 + \frac{u}{K \hbar} \left( \partial_x \phi(x) \right)^2\right], \hspace{.5cm}
\end{eqnarray}

\noindent where the two Tomonaga-Luttinger parameters, $u$ and $K$, depend on the microscopic model. The first, $u$, plays the role of a renormalized ``Fermi velocity,'' which would relate the time and space dimensions of the quantum system. In the polymer system it is unitless and relates the lateral and longitudinal coordinates. The second parameter, $K$, can be treated via a rescaling of the fields [$\phi(x) \rightarrow \phi(x)/\sqrt{K}$, $\theta(x) \rightarrow \sqrt{K} \theta(x)$] which maps the system back onto the $K=1$ case, which corresponds to simply noncrossing systems.
In this way, systems having a wide range of interactions, including those considered in Secs.~\ref{sec:cp} and~\ref{sec:lr}, can be mapped onto the simply noncrossing system of polymers---the very system considered in I. The cost to this rescaling of the fields is that it destroys the relatively straightforward relationship between the fields $\theta(\cdot)$ and $\phi(\cdot)$ [which describe the emergent polymer fluid) and the microscopic degrees of freedom (i.e., the original polymer lines) given in Eq.~(\ref{eq:phitheta}].

Although the bosonized Hamiltonian lacks the inter-polymer length-scale $w/N$ explicitly, the definitions of the fields in Eq.~(\ref{eq:phitheta}) retain the short-distance behavior. From this Hamiltonian, the bosonized action $S$ follows as well:

\begin{eqnarray}
\label{eq:boseaction}
S = \frac{\hbar}{2 \pi K} \int d x \, d \tau \, \left[ \frac{1}{u} \left(\partial_\tau \phi \right)^2 + u  \left(\partial_x \phi \right)^2 \right],
\end{eqnarray}

\noindent associated with which there is a polymer partition function $\mathcal{Z}$, given by

\begin{eqnarray}
\label{eq:bosez}
\mathcal{Z} = \int \mathcal{D} \phi(x,\tau) \hspace{6cm}
\\ \nonumber
\hspace{1cm} \times \exp
\left[
-\frac{1}{2 \pi K} \int d x \, d \tau \, \left( \frac{1}{u} \left(\partial_\tau \phi \right)^2 + u  \left(\partial_x \phi \right)^2 \right)\right].
\end{eqnarray}

\noindent This partition function characterizes the polymer system in terms of the configurations of the field $\phi(x,\tau)$, which is related to polymer density via Eq.~(\ref{eq:phidensity}).
The normalization condition that the system contain $N$ polymers then leads to \emph{periodic} boundary conditions on $\phi(x,\tau)$:

\begin{eqnarray}
\phi(w/2,\tau)-\phi(-w/2,\tau) = 0.
\end{eqnarray}

 Alternatively, we could obtain an action solely in terms of the field $\theta(x,\tau)$, which is related to the polymer slope field. Because of the lack of polymer free ends in the interior of the system, either polymer density or polymer slope fully defines a given configuration of the system. 

The characterization of the polymer system in terms of a harmonic fluid allows one to obtain the polymer correlations. In particular, the density correlations of a large, disorder-free polymer system are given by

\begin{widetext}
\begin{eqnarray}
\label{eq:rhorho}
\big \langle \rho(x+x_0,\tau+\tau_0) \, \rho(x_0,\tau_0) \big \rangle = \left(\frac{N}{w}\right)^2 + \frac{K}{2 \pi^2} \frac{(u \tau)^2-x^2}{\left[x^2+(u \tau)^2\right]^2} +\sum_{m=1}^{\infty}
A_m \frac{\cos(2 \pi m N x /w)}{\left[x^2 +(u \tau)^2\right]^{2 m^2 K}}.
\end{eqnarray}
\end{widetext}

\noindent Note that although the partition function appears to be isotropic, the polymer correlations are in fact sharply anisotropic, owing to the preferred direction of the polymer lines. In particular, the first nonconstant term in Eq.~(\ref{eq:rhorho}) indicates positive (negative) density correlations over longitudinal (lateral) displacements. The other terms, with nonuniversal coefficients $A_m$, describe oscillatory changes in the density correlations over lengths on the scale of the average inter-polymer spacing $w/N$, analogous to Friedel oscillations. When $K<1$, as holds for polymers with long-range repulsion, this oscillatory behavior dominates over long length-scales. For $K>1$, as holds for polymers having a finite contact interaction, the non-oscillatory component dominates instead.

These long-range correlations are characteristic of the disorder-free polymer system. Introducing disorder into the system leads to a characteristic length-scale for correlations, beyond which they are exponentially suppressed (see, e.g., Giamarchi~\cite{giamarchi}).

\section{Topological obstructions}
\label{sec:pin}

In I, we considered the inclusion, into a system of simply noncrossing polymers, of a topological obstruction (a ``pin'') located at $(x_p,\tau_p)$, which some number of polymers $N_L$ passed to the left of, with the remaining polymers passing to its right. The thermal ensemble was restricted to those configurations that met this condition. We showed that the polymer structure around the pin and the free-energy cost of the pin were entirely determined by the polymer correlations encoded in the ground-state wave function. Specifically, the free-energy cost of the pin, which is dominated by the $O(N^2)$ contribution coming from the polymer-polymer interaction, is given by

\begin{eqnarray}
\label{eq:fecost}
\exp \left(-\beta \mathcal{F} \right) = \int_{\mathcal{C}} d \{x_n\} \left| \psi_{\mbox{gs}} \left(\{x_n\} \right) \right|^2,
\end{eqnarray}

\noindent where $\mathcal{C}$ indicates that the integration over polymer coordinates $\{x_n\}$ obeys the topological constraint that exactly $N_L$ polymers pass to the left of the pin:

\begin{eqnarray}
\label{eq:pincon}
-(w/2)<x_{1},\ldots,x_{N_{L}}<x_{p} \hspace{2cm} \nonumber \\
<x_{N_{L}+1},\ldots,x_{N}<(w/2).
\end{eqnarray}

Because the wave function given in Eq.~(\ref{eq:gswavefcncs}) for the polymers with long-range interactions is simply the $\lambda^{\rm th}$ power of the wave function associated with simply noncrossing polymers, it is straightforward to use Eq.~(\ref{eq:fecost}) to show that the dominant density profile $\bar{\rho}(x, N_L, x_p, \lambda)$ on the line $\tau=\tau_p$ is the same for generic values of $\lambda$ (see Sec.~III of I for details of the $\lambda=1$ case):

\begin{eqnarray}
\label{eq:densityprofile}
\bar{\rho}(x,N_L,x_p,\lambda) = \bar{\rho}(x,N_L,x_p,1) = \hspace{2cm}
\vspace{\medskipamount}
\\ \nonumber
\begin{cases}
\displaystyle
\frac{1}{\pwidth}
\sqrt{\frac{\sin(\pi x/\pwidth)-\sin(\pi x_{g}/\pwidth)}
{\sin(\pi x/\pwidth)-\sin(\pi x_{p}/\pwidth)}},&
\begin{matrix}
\!\mbox{for }-\pwidth/2<x<x_{p}\hfill\\
\!\mbox{or }x_{g}<x<\pwidth/2;\hfill
\end{matrix}
\\
\noalign{\medskip}
0,&
\!\mbox{for }x_{p}<x<x_{g} \, .
\end{cases}
\end{eqnarray}

\noindent Note the essential features of this polymer density profile: (i)~a gap, i.e.~a region of zero polymer density (provided density fluctuations are omitted), exists between $x_p$ and some $x_g$ on the rarefied side of the pin; and (ii)~the polymer density diverges immediately to the compressed side of the pin, and remains elevated all the way to the edge of the system (and reduced all the way to the edge on the rarefied side). For simply noncrossing polymers, the entire density profile $\rho(x,\tau)$ was obtained in I, where it was found that the gap has not only a finite width but a finite length in the longitudinal direction. Based on prior studies of the quantum analog~\cite{abanov2,abanov3}---a Calogero-Sutherland system undergoing a large fluctuation---the full density profile should remain qualitatively unchanged by the presence of long range interactions. Furthermore, the free-energy cost of the pin simply scales as $\lambda$:

\begin{eqnarray}
\Delta \mathcal{F}(\lambda) = \lambda \, \Delta \mathcal{F}(\lambda = 1).
\end{eqnarray}

\subsection{Large fluctuations of the bosonized fluid}

We next ask: What is the effect of the pin on the bosonized polymer fluid? More generally, we seek to determine the probability of a given polymer density profile $\bar{\rho}(x,\tau_p)$. Let us impose some value $\bar{\phi}(x)$ of the field $\phi(x,\tau)$ on the line $\tau=\tau_p$, and require that $\phi(x,\tau)$ vanishes (corresponding to uniform polymer density) far from this line. Likewise, $\phi(x,\tau)$ must vanish at $x=\pm w/2$ so that no polymer lines cross the walls of the system.

As we did in applying the wave function formalism, we search for the dominant polymer configuration, i.e., the one that minimizes the free energy, subject to the pin constraint. This requires that $\phi(x,\tau)$ obeys the condition

\begin{eqnarray}
\label{eq:laplace}
\left(u^{-2}\partial_\tau^2 + \partial_x^2\right) \phi(x,\tau) = 0
\end{eqnarray}

\noindent for $\tau \neq \tau_p$. Thus, determining the polymer configuration and free-energy cost associated with the pin is equivalent to solving an electrostatic boundary-value problem, which we proceed to do in Appendix~\ref{sec:fpin}. The result is a free-energy cost $\mathcal{F}_p$ associated with the polymer density profile $\bar{\rho}$, given by

\begin{eqnarray}
\label{eq:fp}
\mathcal{F}_p = -\frac{1}{\beta K } \left(\frac{\pi}{w}\right)^2
\int_{-\frac{w}{2}}^{\frac{w}{2}} \int_{-\frac{w}{2}}^{\frac{w}{2}} d x \, d x' \bar{\rho}(x) \, \bar{\rho}(x') \hspace{0cm}
\nonumber \\
\times \ln \left| \sin \left(\pi x/w\right) - \sin \left(\pi x'/w\right)\right|.
\end{eqnarray}

\noindent Strikingly, this is precisely the expression we found for $\mathcal{F}_p$ for the case of simply noncrossing polymers, scaled by the interaction parameter $1/K$. This result indicates that polymers having a wide class of generic interactions undergo large fluctuations in essentially the manner found for simply noncrossing polymers. Furthermore, the density profile obtained in I and given in Eq.~(\ref{eq:densityprofile}) is found to apply to polymers with generic interactions.

However, there are limits to the validity of the bosonization procedure when considering large fluctuations. Bosonization relies on the linearizing of the spectrum of low-energy excitations, but of course the spectrum is not linear for all excitation wavelengths. If one considers fluctuations in which $\rho(x)$ differs substantially from its equilibrium value $N/w$, such high-energy excitations should lead to behavior that differs from that predicted via bosonization. In particular, we know for simply noncrossing polymers that higher-order terms in the free energy lead to behavior such as a gap of finite area around a pin or barrier. In contrast, the bosonization procedure lacks these terms, and hence predicts that polymer density grows linearly for $\tau>\tau_p$; cf.~Eq.~(\ref{eq:phip}).

Despite this deficiency in the case of the pin, there are particular strong constraints that one can impose on the polymer system for which bosonization remains reliable. For example, bosonization may legitimately be used to describe the polymer ``evolution'' in the longitudinal direction around a constraint requiring $\rho(x,\tau_p) = \big (N+(\delta \rho)x \big )/w$ with $\delta \rho \ll (N/w)$. In such a situation, where the polymer density remains close to its equilibrium value, bosonization is strictly justified.

\section{Concluding Remarks}
\label{sec:conclusion}

In the present Paper, we have described the implications of general interactions on systems of directed polymers in two dimensions. We have found that the type and strength of the interaction modifies the polymer structure and thermodynamics. To a noteworthy extent, the phenomena described in I in the setting of polymers having an infinite contact repulsion extend, at least qualitatively, to polymer systems subject to a much wider range of interactions.

We have used bosonization to describe the universal properties of emergent polymer fluids having generic interactions. We have shown that there are long-range correlations in disorder-free polymer systems, regardless of the range of inter-polymer interactions. Depending on the value of the Tomonaga-Luttinger liquid parameter $K$, the dominant correlations at long distances may or may not include oscillations in the lateral direction.

We have also studied the effect of topological obstructions on the polymer fluid. We have found that the qualitative features of the response are unchanged by the form of the interaction. There is a sharp increase in the density of polymers immediately on the compressed side of a pin, and the increase in polymer density persists over a long distance. On the decompressed side, a two-dimensional gap region opens up; in it the density of polymers is negligible. The restoring force on the pin is proportional to $N^2$, indicating that the dominant response results from interaction effects. Finally, we have described the way in which the density profile, gap region, and restoring force depend on polymer interactions.

\begin{acknowledgments}
We thank Jennifer Curtis, Thierry Giamarchi, Sarang Gopalakrishnan, Michael Pustilnik, Andrew Zangwill and Shina Tan for valuable discussions. One of us (P.M.G.) thanks for its hospitality the Aspen Center for Physics, where part of the work reported here was carried out. This work was supported by Grants No. NSF DMR 09-06780 and DMR 12-07026 (P.M.G.), and by an NDSEG Fellowship (D.Z.R.).
\end{acknowledgments}

\appendix

\section{Short-distance behavior of the polymer system}
\label{sec:ell}

The polymer partition functions that we consider in this Paper contain certain pathologies in their continuum limit that can be controlled via a short-distance regularization. Consider, e.g., the partition function for a single polymer not subject to interactions or external potential, which formally reads

\begin{eqnarray}
\mathcal{Z} = \int \mathcal{D}x(\cdot) \exp \left( -\frac{A \beta}{2} \int d \tau \, \left(\partial_\tau x(\tau)\right)^2 \right).
\end{eqnarray}

\noindent To make sense of this formal object, let us divide the longitudinal coordinate into $M+1$ discrete segments, so that $\mathcal{Z}$ is approximated by

\begin{eqnarray}
\label{eq:zm}
\mathcal{Z}_{(M)} = \int d x_0 \, d x_1 \, \cdots d x_M \hspace{3cm}
\nonumber \\
\exp \left( -\frac{A \beta}{2} \sum_{m=0}^{M} \frac{M}{L} \left(x_{m}-x_{m-1}\right)^2 \right).
\end{eqnarray}

\noindent Note that, here, the $\{x_m\}$ are the coordinates of longitudinally separated segments of a single polymer. Proceeding in the manner of time-slicing for a quantum path integral (see, e.g., Refs.~\cite{kleinert, feynmanhibbs}), one may make use of the following Gaussian integral to eliminate the interior degrees of freedom:

\begin{eqnarray}
\int_{-\infty}^{\infty} d y \, \exp \left( - C \left[ (x-y)^2 + (y-z)^2 \right] \right) = 
\nonumber \\
\sqrt{ \frac{\pi}{2 C} } \exp \left[-C\left(x-z\right)^2/2\right].
\end{eqnarray}

\noindent By using this relationship, we see that the partition function for a single polymer line that travels from $x_0$ to $x_f$ and has $M-1$ interior segments is given by

\begin{eqnarray}
\mathcal{Z}_{(M)} = \frac{1}{\sqrt{M!}}\left(\frac{2 \pi L}{A \beta M}\right)^{(M-1)/2}
\hspace{2cm}
\nonumber \\
\times
\exp\left[- \frac{A \beta}{2 L} \left(x_f-x_0\right)^2\right].
\end{eqnarray}

\begin{figure}[hh]
\centerline{
\includegraphics[width=.5\textwidth]{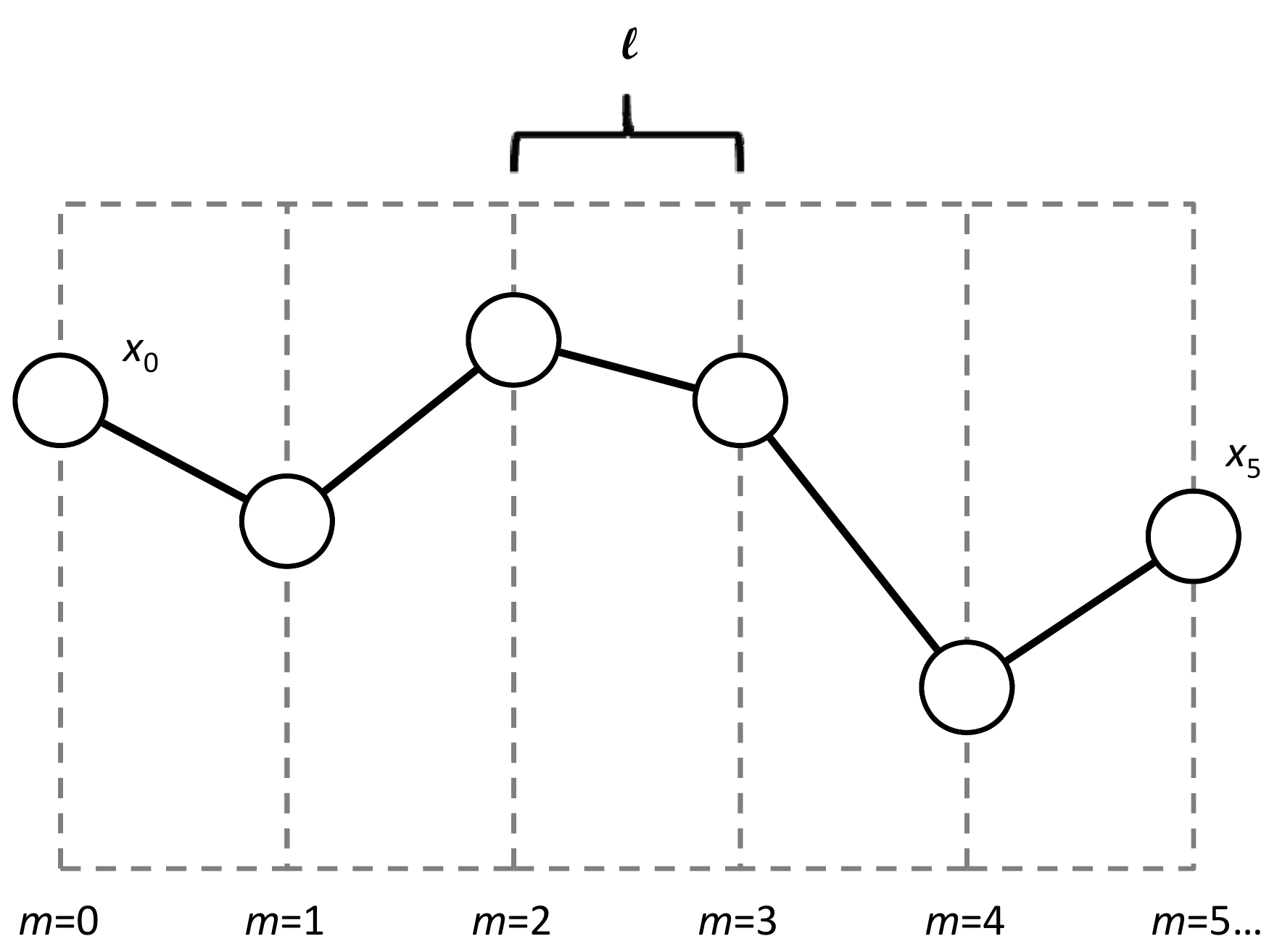}}
\caption{\label{fig:segments}
A configuration of the first five primitive segments of a single polymer. Although the polymer configurations over longer distances do not depend on the cutoff length $\ell$, the root mean squared polymer slope is proportional to $\ell^{-1}$.
}
\label{fig:beaddiagram}
\end{figure}

\noindent We see thus that the free polymer fluctuates with a Gaussian distribution that does not depend on the number of internal segments. The polymer structure over long distances does not depend on this short-distance behavior. 

However, this does not mean that the short-distance behavior can be ignored entirely in the determination of thermodynamic properties of the polymer fluid. Let us define the short-distance cutoff length $\ell \equiv L/M$. Then, the $N$-polymer partition function in Eq.~(\ref{eq:zm}) contains the factor 

\begin{eqnarray}
\mathcal{Z}_{\ell} \equiv \left(\frac{2 \pi \ell}{A \beta}\right)^{NL/2 \ell},
\end{eqnarray}

\noindent which describes the short-distance behavior of the polymers. This factor is not connected to the structure of the polymer lines $\{x_n(\tau)\}$ over distances long compared to $\ell$, but it does influence thermodynamic observables that depend on temperature or the tension parameter. For example, the mean squared deflection $\big \langle \dot{x}(\tau)^2 \big \rangle$
of a free long polymer line is given by

\begin{eqnarray}
\langle \dot{x}(\tau)^2 \rangle = -\frac{2}{\beta L N} \partial_A \ln \mathcal{Z}=\frac{1}{A \beta \ell}.
\end{eqnarray}

\noindent Thus, if we try to take the $\ell \rightarrow 0$ limit we find that short-distance polymer deflections diverge. That is, the more closely we examine an ideal polymer line within the model, the steeper its slope seems to be.  
More generally, the polymer partition function depends on polymer interactions $V(x-x')$ and on one-body potentials $\Phi(x)$ and so the entire partition function for interacting polymers is given by

\begin{eqnarray}
\label{eq:eigfcnexpansion2}
\mathcal{Z}=\mathcal{Z}_\ell \times
(\{x_{n}^{f}\}\vert{{\rm e}^{-\qham\qtime/\hbar}}\vert\{x_{n}^{i}\})
\hspace{3cm}
\\ \nonumber
\qquad\quad = \mathcal{Z}_\ell \times
\sum\nolimits_k e^{-E_k\qtime/\hbar}\,
\psi_k(\{x_{n}^{f}\})\,
\psi_k^\ast (\{x_n^{i}\}).
\end{eqnarray}

\noindent Here, $\{\psi_k(\{x_n\})\}$ and $\{E_k\}$ are the normalized eigenfunctions and eigenvalues of the associated quantum system, as discussed in I. Thus we see that the short-distance cutoff-dependence is isolated in the interaction-independent portion of the partition function $\mathcal{Z}_\ell$ and can be separated from the interaction-dependent physics.

\section{Free-energy cost of rare configurations of bosonized polymers}
\label{sec:fpin}

\begin{figure}[hh]
\centerline{
\includegraphics[width=.5\textwidth]{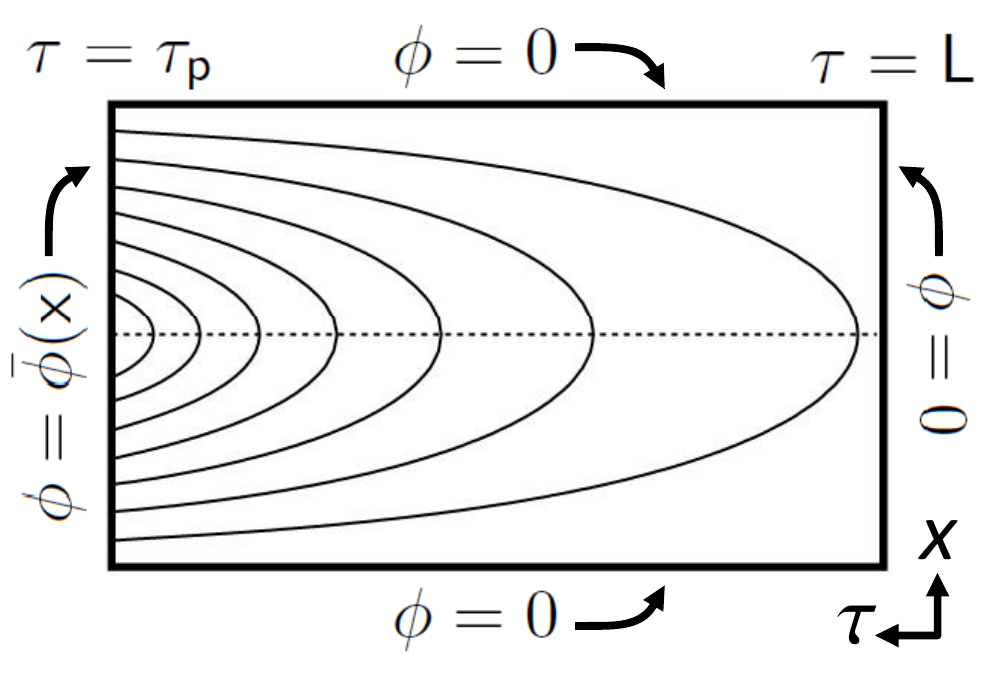}}
\caption{
The bosonized field $\phi(x,\tau)$ vanishes at the walls of the system $x=\pm w/2$ and $\tau=L$ (and at $\tau=0$). A fluctuation in the polymer configuration at $\tau=\tau_p$ leads there to a nonzero value of the field $\phi(x,\tau_p)=\bar{\phi}(x)$. The curved lines are contours of constant $\phi(x,\tau)$ for the configuration that corresponds to 
the boundary condition 
$\rho(x,\tau)= (N/w) + \delta \rho \, \mbox{sign}(x)$.
}
\label{fig:phidiagram}
\end{figure}

We now determine the dominant configuration $\phi_p(x,\tau)$ of the bosonized field  and associated polymer free-energy cost resulting from the general boundary condition $\phi(x,\tau_p)=\bar{\phi}(x)$. The Fourier series solution of the Laplace equation of Eq.~(\ref{eq:laplace}) consistent with this and the other boundary conditions is (choosing for simplicity units so that $w = \pi$)

\begin{subequations}
\begin{eqnarray}
\label{eq:phip}
\phi_p(x,\tau) = \sum_{j=1}^{\infty} a_n \sin \left(j \left[x+\frac{\pi}{2}\right]\right) e^{-j u \left|\tau - \tau_p \right|}; \\
a_j = \frac{2}{\pi} \int_{-\pi/2}^{\pi/2} d x \, \bar{\phi}(x) \sin \left(j \left[x+\frac{\pi}{2}\right]\right).
\end{eqnarray}
\end{subequations}

The polymer free energy $\mathcal{F}_p$ associated with this configuration [which has the same form as the bosonized action of Eq.~(\ref{eq:boseaction})] is given by

\begin{eqnarray}
\mathcal{F}_p = \frac{2}{ \beta K \pi}\int_{-\pi/2}^{\pi/2} \int_{-\pi/2}^{\pi/2} d x \, d x' \, \bar{\phi}(x) \, \bar{\phi}(x')
\hspace{1cm}
\nonumber \\
\times \sum_{j = 1}^{\infty} \, j \sin \left(j \left[x+\frac{\pi}{2}\right]\right) \sin \left(j \left[x'+\frac{\pi}{2}\right]\right).
\end{eqnarray}

\noindent The sum over $j$ may straightforwardly be performed, e.g., by expressing the summand in terms of exponentials and performing the infinite geometric sums. The result is

\begin{eqnarray}
\mathcal{F}_p = -\frac{1}{\beta K \pi} \int_{-\pi/2}^{\pi/2} \int_{-\pi/2}^{\pi/2} d x \, d x' \, \bar{\phi}(x) \, \bar{\phi}(x')
\nonumber \\
\times \frac{\cos  x \, \cos  x' }{\left(\sin x-\sin x' \right)^2}.
\end{eqnarray}

Next, we transform the result for the free-energy cost expressed in terms of $\bar{\phi}(x)$ into one in terms of the polymer density $\bar{\rho}(x)$ on the line $\tau=\tau_p$.
In the continuum limit, in which one considers only length-scales greater than the inter-polymer length, one may invoke the bosonization correspondence of Eq.~(\ref{eq:phidensity}), retaining only the $p=0$ term, $[(N/w) - (\partial_x \phi(x)/\pi)] \sim \rho(x)$. Then, via integrating by parts in $x_1$ and $x_2$ separately, we obtain the free-energy cost of the large fluctuation in terms of the polymer density on the line $\tau = \tau_p$ (restoring physical units of length) given in Eq.~(\ref{eq:fp}).


\begin{thebibliography}{99}
\bibitem{tlocal}
In general, an interaction between polymer segments could depend not only on their lateral separation $x_n - x_{n'}$ but also on the longitudinal separation $\tau_1-\tau_2$. We consider here only interactions that are sufficiently short ranged that they may at least be approximated by an interaction that is local in $\tau$.
\bibitem{cutoff}
Strictly speaking, the effective repulsion parameter $c$ depends on the slope of the polymers, which itself depends on the polymer short-distance cutoff parameter $\ell$ as discussed in Appendix~\ref{sec:ell}.
\bibitem{firstlinebose}
Related work effectively invoking bosonization as a technique for addressing the classical statistical physics of directed linelike systems in two dimensions was undertaken for example by Giamarchi and Le Doussal~\cite{doussal} in the context of vortex lines.
\end{thebibliography}
\end{document}